\documentclass[reprint,amsmath,superscriptaddress,amssymb,aps,prb,showkeys,longbibliography]{revtex4-1}
\providecommand{\enquote}[1]{} 
\renewcommand{\enquote}[1]{#1}  
\usepackage{graphicx}
\usepackage{dcolumn}
\usepackage{bm}
\usepackage{color,graphicx,mfirstuc}
\usepackage[colorlinks,linkcolor=blue,
            citecolor=blue,
            hyperindex,
            pdfstartview=FitH,
            plainpages=false]
            {hyperref}
\usepackage{dsfont}

\usepackage{pifont}
\usepackage{comment}
\usepackage{xcolor}
\usepackage{gensymb}

\begin{document}

\preprint{APS/123-QED}

\title{Spin-Orbit Driven Topological Phases in Kagome Materials}\

\author{Chi Wu}
\affiliation{Institute of Theoretical Physics, Chinese Academy of Sciences, Beijing 100190, China}
\affiliation{University of Chinese Academy of Sciences, Beijing 100049, China}

\author{Tiantian Zhang}
\email{ttzhang@itp.ac.cn} 
\affiliation{Institute of Theoretical Physics, Chinese Academy of Sciences, Beijing 100190, China}


\begin{abstract}
Kagome materials have garnered substantial attention owing to their diverse physical phenomena, yet canonical systems such as the $Z_2$-type AV$_3$Sb$_5$ family are often obscured by bulk metallic states, spurring an urgent quest for kagome platforms hosting ideal topological states.
Recently, Zhou et al. proposed the kagome-type IAMX family (IA = alkali metal, M = rare earth metal, and X = carbon group element), which exhibits distinctive ideal topological states; however, their analysis is primarily restricted to the spinless approximation.
In this work, we model relativistic effects in the IAMX family, demonstrating that tuning the spin-orbit coupling (SOC) strength drives topological phase transitions and induces novel topological states, resulting in a rich phase diagram. The configuration of topological surface states evolves continuously as the SOC strength is modulated, consistent with the evolution of the topological phase transition. This suggests a viable route toward designing multi-functional topological devices. First-principles calculations performed on three specific IAMX compounds confirm that SOC governs their topological phases, in complete accord with our model analysis.
\end{abstract}

\maketitle

\section{Introduction}
Topological phase transitions transcend the framework of conventional Landau phase transition theory, and their investigation not only deepens our understanding of the material realm but also forges a nexus between theoretical physics and materials science. To demarcate distinct topological phases, various topological invariants haven been established, from the seminal TKNN number~\cite{thouless1982quantized,PhysRevLett.95.146802}, which characterizes the integer quantum Hall effect, to the mirror Chern number rooted in crystal symmetries~\cite{fu2007topological,fu2011topological,hsieh2012topological}. Furthermore, studies of topological phase transitions have accelerated the discovery of emergent materials such as Weyl and Dirac semimetals~\cite{murakami2007phase,wang2012dirac,lv2015experimental,wan2011topological,lv2015observation,xu2015discovery,wu2018nodal,weng2015weyl,yan2017nodal,gao2018class}.

The introduction of the Haldane model~\cite{haldane1988model} has elevated lattice models with topological characteristics to a pivotal area of research in condensed matter physics~\cite{yoshida2023polarization,PhysRevB.108.075160,kane2005quantum,bark2025stacking,zhang20252d}. Among these, kagome lattice materials stand out due to their distinctive geometric and electronic properties\cite{jiang2023kagome,feng2021chiral,zhou2022anomalous,wilson2024v3sb5}, offering a superior platform for investigating geometric frustration~\cite{depenbrock2012nature,balents2010spin,yang2020giant}, band topology~\cite{kang2020dirac,li2017evidence,sun2011nearly,yu2021concurrence,yin2022topological,guo2009topological,xue2019acoustic,bolens2019topological}, and unconventional electronic orders~\cite{yu2012chiral,jiang2023observation,li2021observation,nie2022charge,chen2021roton,neupert2022charge}. In addition, certain prototypical combinations of topological lattices can provide greater flexibility for tuning electronic properties. For example, two-dimensional honeycomb-kagome(HK) bilayer systems have been theoretically predicted to host nodal-ring semimetal phases\cite{Lu_2017,bark2025stacking}, the valley Hall effect\cite{zhang2023anomalous,shao2023topological}, and the quantum anomalous Hall effect\cite{wang2013quantum,zhang2022quantum,li2021room}, among others, and  such systems have also been realized experimentally\cite{PhysRevB.102.214301,Lu_2017}. However, these studies are largely limited to two-dimensional scenarios.

Recently, three-dimensional materials comprising stacked kagome and honeycomb lattices, collectively termed the IAMX family (IA = alkali metal, M = rare earth metal, and X = carbon group element)~\cite{zhou2024chemical}, have been investigated through high-throughput calculations and tight-binding models. These materials were initially identified as nodal line semimetals~\cite{zhang2019catalogue,fang2016topological,yu2019quadratic,burkov2011topological} in the absence of spin-orbit coupling (nonSOC), displaying unique electronic properties influenced by their chemical compositions. However, relativistic effects (SOC) in certain IAMX compounds are particularly pronounced, exerting a marked influence on their topological properties, whose precise role has yet to be systematically elucidated.

This work is organized as follows. We first present three materials from the IAMX family, i.e., LiYC, LiNdGe, and KLaPb in Sec.~\ref{sec.2}, revealing that variations in the strength of relativistic effects exert a pronounced impact on the energy bands near the Fermi surface. In Secs.~\ref{sec.3a} and \ref{sec.3b}, We  formulate a minimal four-band $k\cdot p$ model derived from the band structures of LiYC, incorporating relativistic effects. By tuning the SOC term in the model, we obtain phase diagrams and track the evolution of surface states.  Finally, in Sec.~\ref{sec.3c}, we examine the topological surface states and spin textures of these compounds based on maximally localized
Wannier functions constructed by using
the Wannier90 code (DFT+TB)\cite{mostofi2008wannier90}, establishing a direct correspondence between surface and bulk topology. Summary and final remarks are provided in Sec.~\ref{sec.4}.

\section{IAMX family: Crystal structure and electronic band structure}\label{sec.2}
\begin{figure*}[t]
\includegraphics[width=0.8\textwidth]{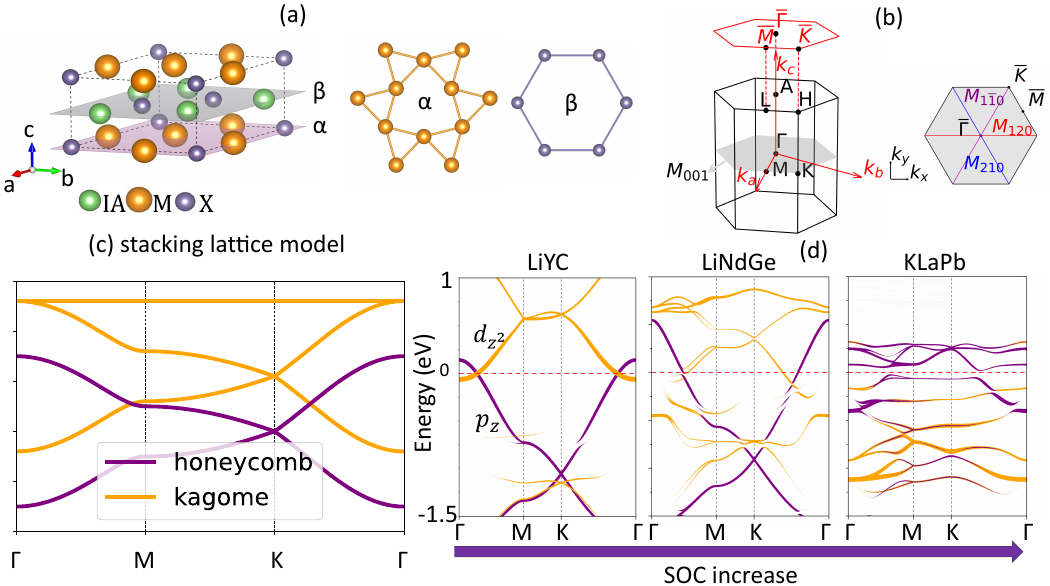}
\caption{\label{fig:crystal} \textbf{(a)} The crystal structure of the IAMX family, where IA, M, and X atoms are denoted by green, yellow, and purple spheres, respectively. The X atoms arrange in a honeycomb lattice configuration within layer $\beta$, whereas the M atoms organize into a distorted kagome lattice structure in layer $\alpha$. \textbf{(b)} The bulk Brillouin zone (BZ) and its projection onto the (001) surface are depicted. The three mirror planes orthogonal to the $\mathrm{M}_{001}$ plane are illustrated on the right. \textbf{(c)}  Schematic energy bands for the system with stacked kagome and honeycomb lattice. \textbf{(d)} The projected band structures (PBSs) in the mirror $k_z=0$ for LiYC, LiNdGe and KLaPb from \textbf{DFT}, wherein the yellow segments represent contributions from the $d_{z^{2}}$ orbitals (kagome layer), and the purple segments indicate the origin from the $p_z$ orbitals (honeycomb layer).}
\end{figure*}

The crystalline structure of the IAMX family exhibits a hexagonal lattice belonging to the space group $P \overline{6}2m$\cite{zhou2024chemical,momma2008vesta,wang2021vaspkit}. As illustrated in Fig. \ref{fig:crystal}a, the IAMX compounds comprise two distinct atomic layers, denoted as $\alpha$ and $\beta$, stacked alternately with interlayer chemical bonding. In layer $\alpha$, X atoms reside at the Wyckoff position $1b$, while M atoms at Wyckoff position $3g$ form a distorted kagome lattice, which can be obtained by rotating the adjacent triangles of a regular kagome lattice about their shared corner atoms
and exhibits characteristic features of the regular kagome lattice, such as flat bands and Dirac points. More details about distorted kagome lattice are in the  Supplemental Material\cite{SM}. Conversely, in layer $\beta$, IA atoms occupy Wyckoff position $3f$, and X atoms at $2c$ establish a honeycomb lattice.
The corresponding first Brillouin zone (BZ) and its [001] projection are displayed in Fig.~\ref{fig:crystal}b. The BZ hosts four mirror planes, a threefold rotation axis $C_{3z}$ along $k_{z}$, and  three twofold rotation axes orthogonal to $C_{3z}$.The mirror plane $M_{001}$ and the three additional perpendicular mirrors play key roles in preserving nodal rings and constraining the spatial distribution of Weyl points in the nonSOC and SOC cases, respectively.

Since the Fermi surface is mainly contributed by the orbitals of the honeycomb lattice and the kagome lattice~\cite{zhou2024chemical}, the essential physics can be captured by a minimal stacked honeycomb-kagome model. Here, we take representations $A_1$ of kagome and $A^{''}$ of honeycomb into account. A schematic band structure without SOC is given in Fig.~\ref{fig:crystal}c. In this case, the kagome-derived bands lie higher in energy than those of the honeycomb, reflecting the higher onsite energy of the M $d$ orbitals compared with the X $p$ orbitals. The disparity in energy levels induces a band crossing near the $\Gamma$ point, engendering a topological band inversion. This inversion signals the presence of the nodal ring phase in the nonSOC case~\cite{zhou2024chemical}.

To further examine how SOC modifies this topological phase, we carried out first-principles calculations on three representative IAMX compounds: LiYC, LiNdGe, and KLaPb. These systems were selected to cover weak, intermediate, and strong SOC regimes, respectively. Their lattice constants are $a$($=b$) = 6.272 Å, 7.312 Å, and 8.799 Å, with a $c/a$ ratio of about 0.58. In the absence of SOC, all three compounds fall into the same class of nodal ring semimetals (NRSMs), characterized by a single nodal ring confined to the $k_z=0$ plane and centered at the $\Gamma$ point~\cite{zhou2024chemical}.
Upon inclusion of  SOC, their band structures around the Fermi energy evolve distinctively, signifying variations in their topological properties, as depicted in Fig.~\ref{fig:crystal}d.
The projected band structures further indicate that the energy bands in proximity to the Fermi surface are predominantly composed of the $d_{z^2}$ orbitals of the M atoms from the kagome lattice and the $p_z$ orbitals of the X atoms from the honeycomb lattice, consistent with Ref.\citenum{zhou2024chemical}. 
Given that the SOC effects are primarily governed by the X element, KLaPb exhibits the strongest SOC influence, whereas the SOC in LiYC is sufficiently weak to be regarded as negligible~\cite{blume1962theory}. This attribute classifies LiYC as a NRSM. 
Compared to the nonSOC case, the energy bands below the Fermi surface in LiNdGe and KLaPb undergo a reordering. However, this band inversion does not induce new topological transitions in these materials.
Comprehensive details of the band structures, density of states (DOS), projected band structures, and irreducible representations (irreps) are provided in the Supplemental Material\cite{SM}.

\begin{figure*}[t]
\includegraphics[width=0.8\textwidth]{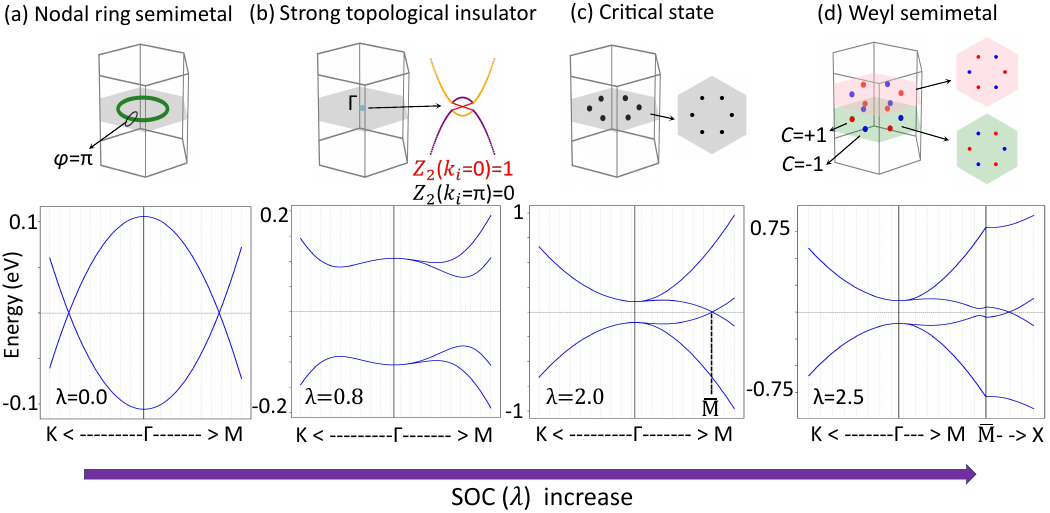}
\caption{\label{fig:bands} The energy bands and topological invariants derived from the minimal four-band $\boldsymbol{k} \cdot \boldsymbol{p}$ model are presented for varying strength of SOC ($\lambda$). From left to right, $\lambda$ increases sequentially from zero. Specific values for $\lambda$ are shown in \textbf{(a)}-\textbf{(d)}, leading to the emergence of distinct topological phases (here $\eta$ is assigned a value of 3). \textbf{(a)} Nodal ring semimetal (NRSM), featuring a single nodal ring (highlighted in green) residing within the $k_z$=0 plane, \textbf{(b)} Topological insulator (TI), \textbf{(c)} The critical state, with the degeneracy points denoted by black dots, and \textbf{(d)} Weyl semimetal (WSM), characterized by six pairs of Weyl points symmetrically positioned on either side of the $k_z$=0 plane.
}
\end{figure*}
\section{Results and discussion}\label{sec.3}

\subsection{\label{subsec:kp}The $k\cdot p$ effective Hamiltonian for IAMX family}\label{sec.3a}

To study the impact of SOC on the topological properties of the IAMX family materials, we develop the minimal low-energy effective models to derive analytical expressions.
Starting with nonSOC, we observe that the two bands that form the nodal ring in LiYC are well-separated from the other bands. Thus, a 2$\times$2 ${k}\cdot {p}$ Hamiltonian can be formulated based on the irreps $\{\Gamma_1,\Gamma_4\}$. The projected band structure of LiYC shows that these two representations predominantly originate from the 3g-Y  $d_{z^2}$ orbitals and the 2c-C $p_z$ orbitals, respectively. To distinguish between them, we therefore directly adopt orbital labels  $\{|d_{z^2}\rangle, |p_z\rangle\}$.
 The nonSOC Hamiltonian, to the leading order in the wave vector $\vec k$, is given by:

\begin{equation}
H_{\text{nonSOC}}(k) = \epsilon(k) +
\begin{pmatrix}
M(k) & A k_{z} \\
-A k_{z} & -M(k)
\end{pmatrix},
\label{eq:one}
\end{equation}
where $\epsilon(k)=C_{0}+C_{1}(k_{x}^2+k_{y}^{2})+C_{2}k_{z}^2$, $M(k)=M_{0}+M_{1}(k_{x}^2+k_{y}^{2})+M_{2}k_{z}^2$, and $A=A_{0}i$. In this context, all parameters are assumed to be real, with the condition $M_{0}M_{1}<0$ to facilitate the band inversion within the mirror plane $M_{001}$. These parameter values can be directly extracted from the wavefunctions obtained via density functional theory (DFT)~\cite{hohenberg1964density,kohn1965self,kresse1993ab,kresse1996efficiency,zhang2023vasp2kp}, as illustrated in the Supplemental Material\cite{SM}.

The eigenvalues of Eq.~(\ref{eq:one}) are expressed as $E_{\pm}=\epsilon(k)\pm \sqrt{M(k)^{2}+(A_{0}k_{z})^2}$, yielding a gapless nodal ring located at $(k_{x}^2+k_{y}^{2}=-{M_{0}}/{M_{1}},k_{z}=0)$. This nodal ring possesses full in-plane rotational symmetry, which may be reduced to $C_{3z}$ symmetry when higher-order $k$ terms or an expanded parameter space are considered. The theoretical framework can be extended to LiNdGe, KLaPb, and other members of the IAMX family.

Accounting for SOC necessitates the employment of double group symmetry for an accurate description. The transformation of irreps at the $\Gamma$ point~\cite{kosterproperties,gao2021irvsp}, corresponding to Eq.~(\ref{eq:one}) and its associated basis functions, is given by:
\begin{gather}
\{\Gamma_1, \Gamma_4\}\otimes \Gamma_7 = \{\Gamma_7, \Gamma_8\},\label{eq:two}\\
\{|d_{z^2},\uparrow\rangle,|d_{z^2},\downarrow\rangle, |p_z,\uparrow\rangle,|p_{z},\downarrow\rangle\}.
\label{eq:three}
\end{gather}

Employing the  basis functions, we initially examine the band degeneracy. Taking the $\Gamma_7$ irrep as a case study, the two-band $k \cdot p$ model can be constructed as:

\begin{equation}
H_{\Gamma_{7}}(k) = \epsilon(k) +
\begin{pmatrix}
D(k) & -Gk_zk_{+}^{2} \\
Gk_zk_{-}^{2} &-D(k)
\end{pmatrix},
\label{eq:four}
\end{equation}
where $D(k)=D_{0}(k_{x}^{2}-\frac{k_y^{2}}{3})k_y$ and $G=G_{0}i, k_{\pm}=k_x \pm ik_y$. The function $\epsilon(k)$ is identical to that presented in Eq.~(\ref{eq:one}), and no differentiation is made here. The eigenvalues of the model are given by:
\begin{equation}
E_{\pm}=\epsilon(k) \pm \sqrt{D(k)^{2}+(G_{0}k_z)^{2}[(k_x^{2}-k_{y}^{2})^{2}+(2k_{x}k_{y})^{2}]}.
\label{eq:five}
\end{equation}

The second term on the right-hand side represents the influence of SOC, and $D(k)$ assumes the Dresselhaus type~\cite{dresselhaus1955spin,jancu2005atomistic}. The degeneracy of $E_{\pm}$ is preserved only in the following cases:
$$
\begin{cases}
  k_z=k_y=0, \qquad\qquad\qquad & {\rm (case.1)}\label{case1}\\
  k_z=0,k_{y}=\pm \sqrt{3}k_x, \qquad\qquad\qquad & {\rm (case.2)}\label{case2}\\
  k_x=k_y=0.\qquad\qquad\qquad & {\rm (case.3)}\label{case3}
\end{cases}
$$
Evidently, band degeneracy is maintained along the high-symmetry directions $\Gamma-A$ and $\Gamma-K$, in agreement with the analysis of irreps and band structures derived from DFT.

We  proceed to construct a minimal model utilizing the basis set ordered as $\{|d_{z^2},\uparrow\rangle, |p_z,\uparrow\rangle, |d_{z^2},\downarrow\rangle, |p_z,\downarrow\rangle\}$.Under such a spin ordering, the SOC can be decomposed into spin-conserving terms along the main diagonal, which take the form of $\sigma_z$ and do not induce spin flipping, and spin-flipping terms in the off-diagonal blocks, which take the form of $\sigma_{x,y}$. Following the notation in Ref.~\citenum{winkler2003spin}, the model can be written explicitly as follows:

\begin{equation}
H_{4\times4}^{soc}(k) = 
\begin{pmatrix}
H_{\text{nonSOC}}(k) & 0 \\
0 & H_{\text{nonSOC}}(k)
\end{pmatrix}+ 
\begin{pmatrix}
H_z^{soc} & H_{xy}^{soc}  \\
H_{xy}^{soc\  \dag} & -H_z^{soc}
\end{pmatrix},
\label{eq:six}
\end{equation}
where the first term on the right-hand side represents a model in the absence of SOC, essentially an extension of Eq. (\ref{eq:one}). The subsequent term corresponds to the SOC effects, which can be obtained by the theory of invariants based on the double group symmetry $D^d_{3h}$. To second order in perturbation theory, the term $H_{z}^{\mathrm{soc}}$ vanishes; whereas
$H_{xy}^{\mathrm{soc}} = i(S_{1}k_{-}\sigma_{x} + S_{2}k_{+}^{2}\sigma_{y})$, a form that incorporates $k_{-}$ and $k_{+}^{2}$ terms and remains invariant under $C_{3z}$ rotational symmetry.

SOC in hexagonal systems has been extensively and systematically studied\cite{kochan2017model,winkler2003spin}. In many representative cases, band inversion and gap closure occur at $K(K^{\prime})$ valleys, which allows one to  expand a tight-binding model around the valleys\cite{xiao2007valley,ren2015single,zhou2019spin}. For example, in graphene under a transverse electric field or in the presence of a substrate, the point-group symmetry is reduced to $C_{6v}$, the $k\cdot p$ model at the valleys of which hosts a constant intrinsic SOC term, as well as linear-$k$ terms corresponding to pseudospin inversion asymmetry (PIA) and Rashba effects\cite{kochan2017model}. The emergence of such constant SOC terms is typically associated with the absence of certain symmetries at the K valleys, such as mirror or time-reversal symmetries. Consequently, SOC constant terms are generally forbidden at the $\Gamma$.
In fact, $k$-dependent SOC terms are more generic in expansions around $\Gamma$. For instance, in the prototypical Kane–Mele model\cite{kane2005quantum,kane2005z}, when expanded around $\Gamma$, the SOC terms naturally appear as $k$-dependent SOC terms in different orders. In Na$_3$Bi and Bi$_2$Se$_3$\cite{wang2012dirac,liu2010model}, the SOC terms are dominated by odd-order terms in $k$. In contrast, for HK lattices and Quadratic DSM Y$_3$XC (X=Tl,Ga)\cite{Lu_2017,tian2020spin}, symmetry constraints prohibit linear SOC terms, while second-order SOC terms remain allowed. Therefore, for 
$k \cdot p$ models constructed around $\Gamma$, the form of SOC depends sensitively on the symmetry and bases.
It should be emphasized that the SOC term in Eq.~(\ref{eq:six}) is obtained directly from a symmetry analysis of the crystal and represents the intrinsic SOC.

\subsection{\label{subsec:transition}Topological phase diagrams based on the $k\cdot p$ model}\label{sec.3b}

Our application of L\"{o}wdin perturbation theory~\cite{lowdin1951note,voon2009kp,kane1966k,liu2010model,wu2020higher} reveals that the $k_{-}$ term arises from the first-order SOC contribution, while the $k_{+}^{2}$ term originates from the second-order SOC effect~\cite{rothe2010fingerprint,bernevig2006quantum,faria2016realistic}. Detailed derivations are in the Supplemental Material\cite{SM}.
From the standard form of the SOC Hamiltonian, $H^{soc}=\frac{\hbar}{4m_{0}^{2}c^{2}}(\vec \sigma \times \nabla V)\cdot \vec p$, it follows that in an ideal system where the Fermi surface involves only a few bands of interest, enhanced SOC strength amplifies both $S_{1}$ and $S_{2}$. This relationship becomes analytically tractable under the assumption that modifications to the SOC originate from either: (i) a correction to the velocity, $c \to v$ , or
(ii) a scalar potential scaling, $V \to \gamma\cdot V$. These mechanisms yield 
\begin{equation}
\left\{
\begin{aligned}
S_{1} &= \lambda \text{\ and} \\ 
S_{2} &= \eta \cdot \lambda^{2},
\end{aligned}
\right.
\label{eq:seven}
\end{equation}
where $\lambda$ determines the strength of the SOC and $\eta$ remains a constant that is material-dependent. 
By varying the parameter $\lambda$ in Eq.~\eqref{eq:seven} while holding $\eta$ constant, a series of topological phase transitions can be realized within the model, as illustrated in Fig.~\ref{fig:bands}.

Fig.~\ref{fig:bands} illustrates that as the parameter $\lambda$ increases from zero, the system undergoes a series of topological phase transitions, marked by the repeated opening and closing of the band gap. The system transitions from a nodal-ring semimetal (NRSM) to a strong topological insulator (sTI), eventually reaching a critical point where the band gap closes completely.
Two of the critical points are distributed on the intersections of $k_z$=0 and $k_x$=0($k_y > 0$), with eigenvalues of $E(k_x=k_z=0)=\pm \sqrt{(M_{0}+M_{1}k_y^{2})^{2}+k_y^{2}(S_{1}\pm S_{2}k_{y})^{2}}$, i.e., the gapless critical state emerges under the condition $S_{2}=\pm S_{1} \sqrt{-M_{1}/M_{0}}$. Ultimately, the system transforms into a Weyl semimetal (WSM), with Weyl points symmetrically distributed relative to the $M_{001}$ plane.
\begin{figure*}[t]
\includegraphics[width=0.8\textwidth]{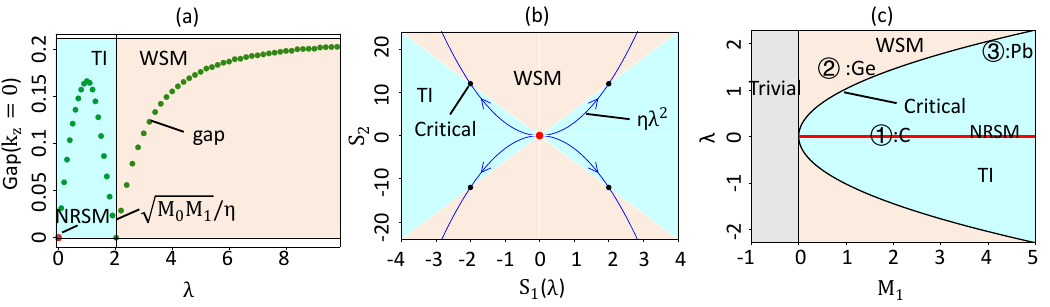}
\caption{\label{fig:Gap} Phase diagrams for the minimal $\boldsymbol{k} \cdot \boldsymbol{p}$ model. \textbf{(a)} Topological phase diagram as a function of the energy gap on the $k_z=0$ plane and the strength of SOC ($\lambda$). The gap closure serves as the signature of the topological phase transition. \textbf{(b)} The phase diagram as a function of $S_1$ and $S_2$ parameters, which corresponds to the first- and second-order SOC term, respectively. The curve represented by $\eta \lambda^{2}$ initiates at zero and traverses both the Weyl semimetal (WSM) and topological insulator (TI) phases (here $\eta$=3). \textbf{(c)} Topological phase as a function of SOC ($\lambda$) and $M_1$. Following the relationship in Eq.~(\ref{eq:seven}), the approximate parameter and phase location for LiYC, LiNdGe, and KLaPb are denoted with \fontsize{11}{0}\selectfont\raisebox{-0.3ex}{\ding{172}}\raisebox{-0.3ex}{\ding{173}}\raisebox{-0.3ex}{\ding{174}}.} 
\end{figure*}

Based on the analysis in Fig.~\ref{fig:bands}, which shows that band gap closing and reopening occur specifically in the $k_z = 0$ plane in conjunction with topological phase transitions, we further explore this behavior in Fig.~\ref{fig:Gap}a. This figure presents the topological phase diagram obtained from Eq.~\eqref{eq:six}, illustrating the relationship between the SOC strength $\lambda$ and the energy gap at the $k_z = 0$ plane (denoted as ``Gap ($k_z$=0)'').

As shown in Fig.~\ref{fig:Gap}a, the system evolves with increasing $\lambda$: beginning as a nodal-ring semimetal (NRSM, red dot), it transitions into a strong topological insulator (sTI), passes through a gapless critical point near $\lambda = \sqrt{M_0 / M_1} / \eta$, and finally enters a Weyl semimetal (WSM) phase upon further increase of $\lambda$, where the gap reopens.

In Fig.~\ref{fig:Gap}b, where $S_1$ and $S_2$ are treated as independent parameters, the phase diagram reveals critical transitions along the boundaries $S_2 = \pm S_1 \sqrt{-M_1 / M_0}$, consistent with earlier discussion. Here, $M_1$ controls the effective mass perpendicular to the $k_z$-direction. In IAMX materials, $M_1$ is typically smaller than $M_2$ due to stronger intralayer hopping—a consequence of short lattice constants and pronounced electronegativity differences—resulting in a highly anisotropic band structure.

\begin{figure*}[t]
\includegraphics[width=0.8\textwidth]{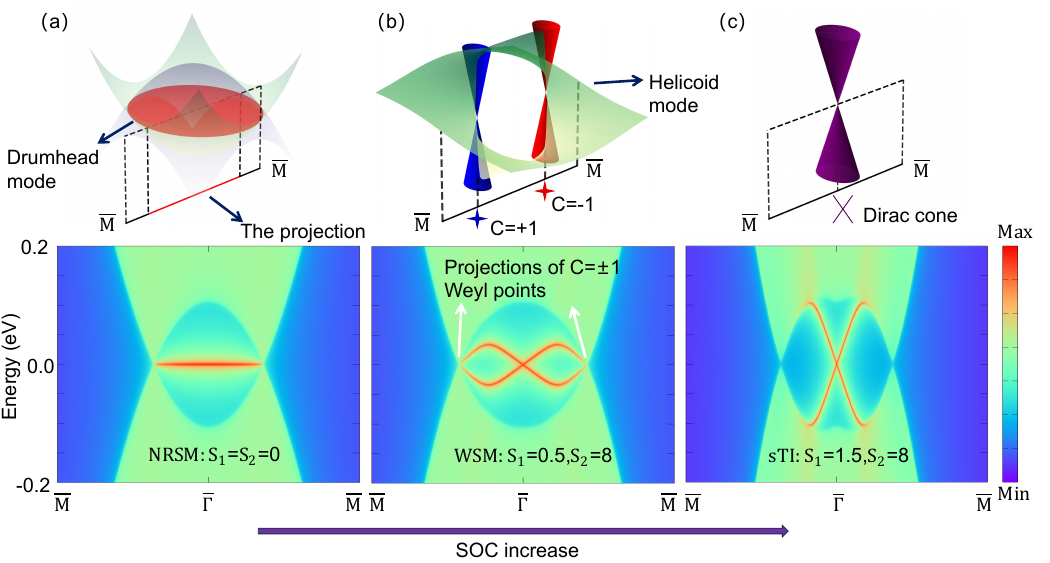}
\caption{\label{fig:kp_ss}The projected surface
states on the (001) surface BZ along $\overline{M}-\overline{\Gamma}-\overline{M}$ for the minimal $\boldsymbol{k} \cdot \boldsymbol{p}$ model, with the top row showing 3D schematics of the surface states for each topological phase (the vertical dimension representing energy).
These three states correspond to distinct combinations of $S_{1}$ and $S_{2}$, demonstrating the evolution of surface electronic structure under varying SOC strength. \textbf{(a)} Surface states in the absence of SOC, corresponding to the NRSM phase. \textbf{(b)} Surface states in the WSM phase. The upper portion shows the helicoid structure associated with a pair of Weyl points of opposite chirality, representing the fundamental building block of surface states in a WSM.
\textbf{(c)} Surface states in the sTI phase, realized with larger  $S_{1}$. }
\end{figure*}

Fig.~\ref{fig:Gap}c shows the topological phase diagram as a function of $\lambda$ and $M_1$, consistent with the relation given in Eq.~\eqref{eq:seven}. The critical condition extends into a black line, reflecting broad tunability of topological phases through these parameters. In band-inversion scenarios, $\lambda$ serves as the primary tuning parameter, underscoring the versatility of the IAMX family for investigating topological transitions via chemical doping. Fitting $k\cdot p$ model to the DFT bands confirm that the compounds LiYC, LiNdGe, and KLaPb fall within this phase diagram, as indicated by the markers in Fig.~\ref{fig:Gap}c.

The phase diagrams in Figs.~\ref{fig:Gap}a and c are based on the idealized theoretical model of Eq.~\eqref{eq:seven}, while real materials may exhibit deviations due to additional perturbations. In contrast, Fig.~\ref{fig:Gap}b possesses broader applicability, serving as a general guide for tuning $S_{1}$ and $S_{2}$ to track the evolution of surface states across phase transitions, as shown in Fig.~\ref{fig:kp_ss}.
In the absence of SOC ($S_1=S_2=0$), Fig.~\ref{fig:kp_ss}a displays the surface states of a NRSM. Introducing finite SOC with a relatively small $S_{1}$ partially lifts the degeneracy of these surface states, except at the $\overline{\Gamma}$ point, and leads to the emergence of a helicoid structure. Fig.~\ref{fig:kp_ss}b shows the surface states in the WSM phase, where the upper panel reveals helical modes originating from a pair of Weyl nodes with opposite chirality, analogous to the Riemann surface of a holomorphic function~\cite{fangtopological}. As $S_1$ increases further, the energy gap fully opens, resulting in an isolated Dirac cone at the $\overline{\Gamma}$ point, shown in Fig.~\ref{fig:kp_ss}c.

\begin{figure*}[t]
\includegraphics[width=0.8\textwidth]{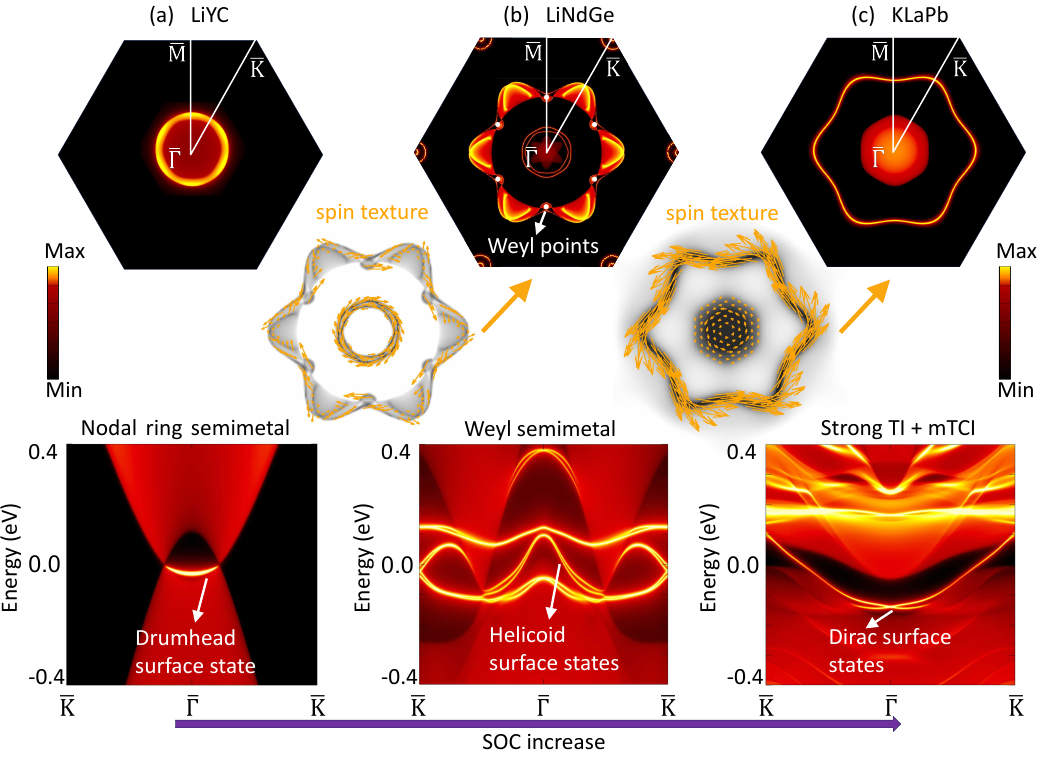}
\caption{\label{fig:SSs} The projected surface states along [001] direction and their corresponding Fermi surfaces for three of the IAMX family materials from \textbf{DFT+TB}. In these figures, the bulk states are depicted in black and dark red, whereas the surface states are highlighted in gold. The surface spin texture for LiNdGe and KLaPb are depicted with orange arrows.  \textbf{(a)} The Fermi surface and drumhead surface state for LiYC. \textbf{(b)} The Fermi arcs and helicoid surface states for LiNdGe, wherein each white dot represent a pair of Weyl fermions with opposite chirality. \textbf{(c)} The Fermi surface and Dirac surface states for KLaPb. }
\end{figure*}

\subsection{\label{subsec:surface states}SOC-enriched topological phases in the IAMX family}\label{sec.3c}

As briefly outlined above, the evolution of surface states is driven by SOC and discussed by the effective model. We now examine this phenomenon in detail for specific materials based on DFT+TB method. It should be noted that the construction and discussion of the effective model above are based on the Fermi-surface properties of LiYC. For LiNdGe and KLaPb discussed below, the Fermi surfaces are not as cleanly composed of $d_{z^2}$ and $p_z$ orbitals. However, the band representation forming the band inversion remains unchanged, as can be verified from the nonSOC projected band structures shown in the Supplemental Material\cite{SM}. It is therefore reasonable to extend the same model description of the phase transition to these two materials.
According to the ``bulk–boundary correspondence''~\cite{chiu2016classification}, the topological nature of the bulk determines the properties of the surface states. Here, we study the effect of SOC on the topological surface states of the three previously mentioned materials, focusing on their (001) surfaces, a $C_{3z}$-symmetric plane containing three vertical mirror planes that protect the symmetry, enforced topological states.

We begin with the WSM LiNdGe, characterized by a symmetry-based indicator of $\mathbb{Z}_3 \times \mathbb{Z}_3 = (1,\ 0)$\cite{po2017symmetry,he2019symtopo,bradlyn2017topological,zhang2019catalogue}. On the (001) surface BZ, six pairs of Weyl points are projected, with each projection (white dots in Fig.~\ref{fig:SSs}b) corresponding to a pair of Weyl nodes carrying opposite chirality.
The Fermi arc configuration exhibits a pairwise connection among every three projected points, enforced by the $C_{3z}$ rotational symmetry. As previously reported for LiNdGe~\cite{zhou2024chemical}, this connectivity is dictated by mirror Chern numbers, each equal to unity, associated with the three vertical mirror planes. As a result, the Fermi arcs remain degenerate along the projection lines of these mirror-invariant planes. A comparison of the surface spin texture with the corresponding surface states further reveals spin–momentum locking within the surface Dirac cone: the conduction branch displays left-handed helicity, whereas the valence branch exhibits right-handed helicity.

In LiNdGe, reducing the SOC to zero, as in the LiYC case, induces a transition from the WSM phase to a NRSM with a Berry phase of $\pi$, as shown in Fig.~\ref{fig:bands}. Any $k$-path intersecting the nodal ring yields a one-dimensional band structure equivalent to that of a Weyl pair, resulting in a surface Fermi arc that connects them, as depicted in Fig.~\ref{fig:SSs}a. Due to the infinite number of such intersecting paths, these arcs coalesce into drumhead-like surface states.

We trace the evolution of the topological surface states from the WSM phase (Fig.~\ref{fig:SSs}b) to the NRSM phase (Fig.~\ref{fig:SSs}a). In the WSM, two helicoid surface states develop along the $\bar{K}-\bar{\Gamma}-\bar{K}$ path, becoming doubly degenerate at $\bar{\Gamma}$. As the SOC strength decreases, the bulk gap along $\bar{K}-\bar{\Gamma}$ narrows, progressively compressing the helicoids. When SOC is fully suppressed, the gap closes completely, and the two helicoids merge into the degenerate drumhead-like surface state characteristic of the NRSM\cite{zhang2021unique}.

Conversely, increasing the SOC strength drives a transition from the WSM phase to a sTI, as shown in Fig.~\ref{fig:SSs}c and illustrated by the evolution from Fig.~\ref{fig:kp_ss}b to Fig.~\ref{fig:kp_ss}c. This transition is marked by the emergence of a single surface Dirac cone at the $\bar{\Gamma}$ point. Symmetry analysis and DFT calculations identify KLaPb as having a symmetry-based indicator $\mathbb{Z}_3 \times \mathbb{Z}_3 = (1,\ 0)$. Using quantitative mapping methods~\cite{song2018quantitative}, the corresponding topological crystalline insulator (TCI) indicators are determined and summarized in Table~\ref{tab:SI}.

\begin{table}[t]
  \caption{\label{tab:SI}The SI and TCI invariants for KLaPb, the definitions of which can be found in the supplementary in Ref.\citenum{song2018quantitative}}
  \begin{tabular}{ccccccc}
    \hline
    $\mathbb{Z}_{3,3}$&weak&$m_{(3)}^{001}$&$m_{(2)}^{1\overline{1}0}$&$g_{\frac{1}{2}\frac{1}{2}0}^{1\overline{1}0}$&$2^{010}$&$2_{1}^{010}$\\
    \hline 
    1,0& 000 & $\overline{2}0$
    &2&$\mathds{1}$ & $\mathds{1}$ &$\mathds{1}$\\
    \hline
  \end{tabular}
\end{table}

Increasing SOC strength drives a progressive enlargement of the bulk energy gap along $\bar{K}$–$\bar{\Gamma}$, accompanied by spatial separation of the helicoid surface states. In KLaPb, the full opening of the bulk gap results in the merger of these helicoids into a single surface Dirac cone, characteristic of a strong topological insulator (sTI), where the conduction branch exhibits right-handed helicity. Furthermore, KLaPb emerges as a promising topological crystalline insulator (TCI), hosting mirror-protected and hourglass topological states inherited from LiNdGe~\cite{song2018quantitative}.

\section{Conclusion}\label{sec.4}

We perform a systematic symmetry analysis of the IAMX family, comprising stacked kagome–honeycomb lattices, by examining the irreducible representations and orbital compositions of the electronic bands in three representative materials, each residing in a distinct topological phase. Using the $\Gamma$-point irreps, we construct two-band (without SOC) and four-band (with SOC) models to elucidate how SOC strength serves as a continuous tuning parameter driving topological phase transitions, ultimately yielding a global topological phase diagram for this kagome system. Guided by this diagram, we simulate and analyze the evolution of surface states using the effective model. Subsequently, we calculate surface states of the three IAMX materials with significantly different SOC strengths using a combined DFT+TB method. All three materials exhibit characteristic topological surface features that can be continuously interconnected via SOC tuning, which  is consistent with the results of the $k\cdot p$ simulations. Our results provide a comprehensive understanding of topological phases in the IAMX family and offer practical guidelines for controlling these phases through doping or isotope substitution.

\begin{acknowledgments}
T. Zhang and C. Wu acknowledge the support from National Key R\&D Young Scientist Project (Grant Nos. 2023YFA1407400 and 2024YFA1400036), and the National Natural Science Foundation of China (Grant Nos. 12374165 and 12047503).

\end{acknowledgments}

\nocite{*}

\newpage
\bibliography{ref}

\end{document}